\documentclass[showpacs, nobalancelastpage, nofootinbib]{revtex4-1}
\usepackage{mathrsfs}
\usepackage{amsmath}
\usepackage{dsfont}
\usepackage{tensor}
\usepackage{amssymb}
\usepackage{graphicx}
\usepackage{yfonts}
\usepackage{mathtools}
\usepackage{wasysym}
\usepackage{siunitx}

\newcommand{\xo}[2]{\overset{#1}{#2}}
\newcommand{\mytitle}[0]{A Smolin-like branching multiverse from multiscalar-tensor theory}
\newcommand{\myauthor}[0]{K.~A.~S.~Croker}
\newcommand{\rmd}[0]{\mathrm{d}}

\usepackage[pdfauthor={\myauthor},pdftitle={\mytitle}]{hyperref}

\begin{document}
\title{\mytitle}
\author{\myauthor}
\email{kcroker@phys.hawaii.edu}
\affiliation{Department of Physics and Astronomy, University of Hawai'i at M\=anoa, 2505 Correa Rd., Honolulu, Hawai`i 96822, USA}
\affiliation{Teoreetilise F\"u\"usika Labor, F\"u\"usika Instituut, Tartu
  \"Ulikool, Ravila 14c, EE-50411 Tartu, Estonia}
\date{\today}

\begin{abstract}
We implement a Smolin-like branching multiverse through a directed,
acyclic graph of $N$ metrics.  Our gravitational and matter actions
are indistinguishable from $N$ decoupled statements of General
Relativity, if one varies with respect to metric degrees of freedom.
We replace $N-1$ metrics with scalar fields by conformally relating
each metric to its unique graph predecessor. Varying with respect to
the $N-1$ scalar fields gives a multiscalar-tensor model which
naturally features dark matter candidates.  Building atop an argument
of Chapline and Laughlin, branching is accomplished with the emergence
of order parameters during gravitational collapse: we bootstrap a
suitably defined $N$ scalar field model with initial data from an
$N-1$ field model.  We focus on the nearest-neighbour approximation,
determine conditions for dynamical stability, and compute the
equations of motion.  The model features a novel screening property
where the scalar fields actively adjust to decouple themselves from
the stress, oscillating about the requisite values.  In the Newtonian
limit, these background values for the scalar fields exactly reproduce
Newton's law of gravitation.
\end{abstract}

 % 04.20.Fy = Lagrangians, variational stuff GR, 04.50.Kd = modified theory of gravity, 95.35.+d = Dark Matter
\maketitle
\section{Introduction}
In 1992, Lee Smolin posed the question\cite{smolin1992did}, ``Did the
Universe Evolve?''  His use of the word ``evolve'' was not in the
physicists' sense of dynamical time evolution, but in the biologists'
sense of speciation.  Smolin proposed that gravitational collapse
produces an offspring Universe with slightly different Standard Model
(SM) parameters from the progenitor: a microphysical analogy to the
microbiology of inheritance.  After many generations, he argued, we
should find ourselves at a local maximum over SM parameter space for
the formation of collapsed objects.  In this way, he proposed a
fascinating resolution to the Hierarchy Problem: it is our notion of
naturalness that we must adjust.

Of the criticisms (e.g. \cite{smolin2006status}) levelled against
Smolin's 1992 proposal, the only one relevant is that it
disagrees with observation (e.g. \cite{rothman1993smolin}).  We regard
this result as refutation only of Smolin's particular fitness
function: reproductive rate.  Note that this fitness function is
appropriate in the simplest of scenarios: e.g. bacteria with unlimited
food.  In any realistic setting, one cannot consider the organism in
isolation; competition and environmental pressures ultimately
determine the fate of \emph{populations}.  In fact, it is observation
of how interacting organisms within populations change over timescales
much greater than an individual's lifespan which led to Darwin's
conclusions\cite{darwin1872origin} on fitness, not the other way
around.

The purpose of this paper is to develop a notion of population and
ancestry via an extension to the Einstein-Hilbert action.  We restrict
ourselves to SM clones for simplicity, as we have yet no
\emph{experimental} basis for speculation on the microphysical origins
of speciation.  Our approach combines two well-motivated lines of
extension to General Relativity (GR): multiscalar-tensor theories and
multiple metric theories.  There is a vast and mature literature on
scalar-tensor theories.  Not only have their observational signatures
been rigorously characterised(e.g. \cite{damour1992tensor,
  damour1993nonperturbative}), but scalar fields are well known to
contribute negative pressure, and are a mainstay of primordial
inflation scenarios\cite{bassett2006inflation}.  Multiple metric
theories also have a long history (e.g. \cite{rosen1963flat}), with
well-known pathologies\cite{TEGP, boulanger2001inconsistency} having
motivated novel and potentially viable
(e.g. \cite{hassan2012bimetric}) approaches.  The result of our effort
will be a classical field theory which makes quantitative predictions
amenable to immediate confrontation with existing and future
astrophysical data.

The rest of this paper is organised as follows.  We first review the
necessary technical background and emphasise some common caveats when
working in multiple metric settings. We then motivate and introduce
our gravitational and matter actions, featuring $N$ metric degrees of
freedom.  Through conformal relations, we establish a consistent
notion of causality across the $N$ metrics by replacing $N-1$ metric
degrees of freedom with scalar degrees of freedom.  We perturbatively
expand the resulting gravitational action, which takes the form of a
multiscalar-tensor theory in the Jordan frame, and determine
conditions for dynamical stability.  We compute the equations of
motion, demonstrate the consistency of our branching process, and
examine the Newtonian limit.  We then discuss very many future
directions and conclude.  An appendix presents a conceptually
related model as a foil to emphasise our particular approach.  A
second appendix outlines an alternate route to a model with many of
the features developed in the main body, with differing aesthetic
criteria that may appeal to some researchers.  Throughout the paper,
we employ units where $c \equiv 1$.  Note that we will define notation
at its introduction, and will often abbreviate ``stress-energy'' as
simply ``stress.''  If context should require distinction, it will be
made explicitly.

\section{The phylogenic model}
We briefly review some of the mathematics involved when considering
multiple metrics defined on a common differentiable manifold
$\mathscr{M}$ of dimension $n$.  First recall any suitable almost
everywhere $C^\infty$ differentiable manifold admits the notion of
curves independently of any geometry.  These objects are maps $C(\tau)
: \mathscr{M} \leftarrow \tau \in [a,b] \subset \mathbb{R}$ which,
when composed with the coordinate charts $\xi^i : \mathbb{R}
\leftarrow \mathscr{M}$, become real-valued differentiable functions
$\xi^i \circ C$.  Derivatives of these compositions can then be used,
in a natural way\cite{o1983semi}, to construct tangent and cotangent
spaces on $\mathscr{M}$.

At one's discretion, notions like metric, connection, and volume form
(Jacobian) can be introduced to augment $\mathscr{M}$.  There is no
natural notion of uniqueness for any of these objects.  Given a
metric, however, one may uniquely construct a torsion-free connection
compatible with this metric.  This Levi-Civita connection is the only
type which we consider below.

Observers using distinct metrics will arrive at distinct physical
conclusions.  For example, fix two events $x^\mu$ and ${x'}^\mu$ on
$\mathscr{M}$ and fix some curve $C$ such that $C(\tau_i)\equiv x^\mu$
and $C(\tau_f)\equiv {x'}^\mu$.  As stated above, this curve has a
field of velocity vectors $\rmd C^\mu/\rmd\tau$ that exist
independently of any geometry.  If $C$ is always time-like with
respect to metrics labelled $p$ and $q$, then we may consider the
proper separation as perceived by observers using metric $p$ versus
observers using metric $q$.  It may happen that
\begin{align}
\int_C \sqrt{\xo{p}{g}_{\mu\nu}\frac{\rmd x^\mu}{\rmd\tau}\frac{\rmd x^\nu}{\rmd\tau}}~\rmd\tau \neq \int_C \sqrt{\xo{q}{g}_{\mu\nu}\frac{\rmd x^\mu}{\rmd\tau}\frac{\rmd x^\nu}{\rmd\tau}}~\rmd\tau .
\label{eqn:proper_times}
\end{align}
Physically, the presence of multiple metrics requires an additional
means of assigning ``measurement apparatus'' to observers, and this
assignment partitions observers into distinct classes. For clarity,
``observers'' continues to mean ``frames which arrive at the same
physical conclusions under diffeomorphism.''  We regard this property
as a virtue of multiple metric theories, in that such theories
simultaneously acknowledge the mathematical non-uniqueness of metric
and remove the unique \emph{class} of observers from GR.  This
uniqueness in GR infamously led to Fock's somewhat amusing complaint
of, ``a widespread misinterpretation of the Einsteinian Gravitation
Theory as some kind of general relativity.''\cite{fock2015theory}.

In the multimetric setting, care must be exercised when executing
common tensor operations.  Raising and lowering indices on an
$q$-labelled object can only be accomplished with the $q$ metric.  For
example
\begin{align}
\xo{q}{R_{\mu\nu}}\xo{p}{g^{\mu\rho}} \neq \xo{q}{R^\rho}_\nu
\end{align}
and cannot be simplified further, without additional information relating
the $p$-metric to the $q$-metric.  Likewise there are now distinct
covariant derivatives, which are only compatible with their paired
metrics.  In other words
\begin{align}
\xo{p}{\nabla}_\mu\left(\xo{q}{g^{\mu\nu}}\xo{p}{T_{\nu\rho}}\right) \neq \xo{q}{g^{\mu\nu}}\xo{p}{\nabla}_\mu\xo{p}{T_{\nu\rho}}
\end{align}
because the $q$ metric is not compatible with the $p$ connection
and so cannot be commuted through this covariant differentiation.

\subsection{The phylogenic matter and gravitational actions}
\begin{figure}
\caption{Directed, acyclic graph representation of couplings,
  conformal relations, and relative orders for a model anchored at
  $p$.  Note terms $O(\kappa^2)$ are discarded in the
  nearest-neighbour model of \S\ref{sec:nearest_neighbor_model}.}
\includegraphics{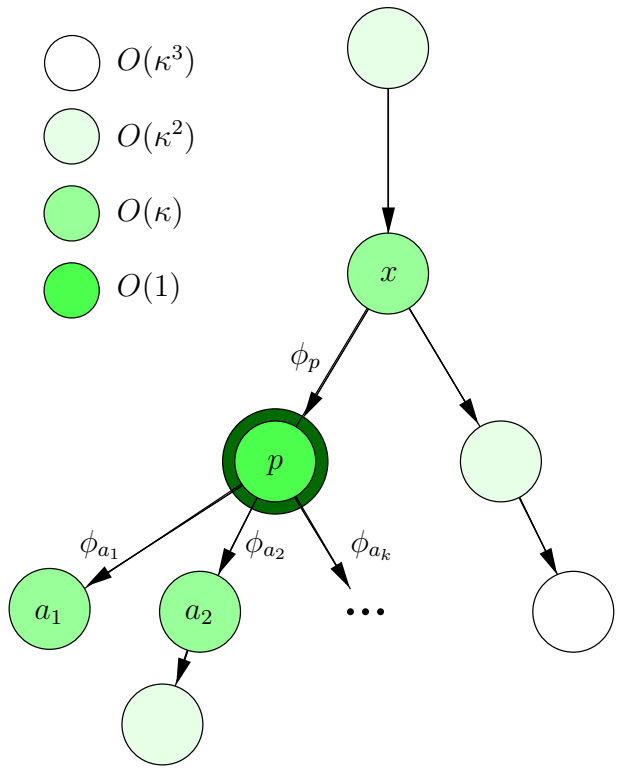}
\label{fig:tree_schematic}
\end{figure}

The utility of GR and its metric foundation, as a quantitative model
for reality, has been rigorously verified over many decades of
spatiotemporal scale.  Consequently, any departures from GR must
either be very slight, subtly hidden, or both.  Consider a population
of $N$ metrics and $N$ copies of the SM.  Introduce a dimensionless
coupling constant $\kappa > 0$, and consider a directed, acyclic
graph (DAG) e.g. the rooted tree of Figure~\ref{fig:tree_schematic}.
Associate to each vertex $q$ of this tree a distinct metric
$g^q_{\mu\nu}$ and stress $T^q_{\mu\nu}$.  Given any two vertices $p$
and $q$ of this graph, let $r(p,q) \geq 0$ be the graph distance
between vertex $p$ and vertex $q$.  This distance is well-defined
because there is a unique simple path between any two vertices of a
tree.  We assign observers to metrics by augmenting a straightforward
generalisation of the canonical matter action, due to Hohmann and
Wohlfarth\cite{hohmann2010repulsive}, with the graph distance between
$p$ and $q$
\begin{align}
\delta \xo{p}{S}_M \equiv -{1\over 2}\int\rmd^4\xi~ \sum_{q}^N  \kappa^{r(p,q)}\xo{q}{T}_{\mu\nu}\delta\xo{q}{g^{\mu\nu}}\sqrt{q}.
\label{eqn:matter_actions}
\end{align}
Here $\sqrt{q}$ is shorthand for the natural volume form induced by
the $q$ metric.  Note that these $N$ matter actions, indexed by $p$,
each enforce\cite{fock2015theory} $N$ distinct statements of the
Equivalence Principle: observers using the $q$ metric will observe $q$
stress to follow $q$ metric geodesics.

Multiple metric gravitational actions are tightly constrained.  With
the exception of Hassan and Rosen's multiple metric
theory\cite{hassan2012bimetric}, Theorem 1.1 of
Boulanger~\emph{et. al.}  \cite{boulanger2001inconsistency} demands
that all metrics must be decoupled.  We thus consider the class of
models indexed by $p$
\begin{align}
\xo{p}{S}_G \equiv \frac{1}{16\pi G}\int\rmd^4\xi~ \sum_{q}^N  \kappa^{r(p,q)} \xo{q}{R}\sqrt{q}
\label{eqn:grav_actions}
\end{align}
where powers of $\kappa$ enter in the same manner as (\ref{eqn:matter_actions}).
Note that these actions suggest a natural perturbative treatment in
powers of $\kappa$.  If one varies the gravitational actions
(\ref{eqn:grav_actions}) with respect to the metric degrees of
freedom, combination with (\ref{eqn:matter_actions}) recovers $N$
distinct statements of \emph{exactly} Einstein's equations for each
$q$.  Thus, there is no way to experimentally distinguish any of these
actions from that of Hilbert and we have successfully hidden our new
degrees of freedom.

\subsection{Conformal relations and ancestry}
\label{sec:ancestry}
An undesirable feature of any model with $N$ possibly distinct metrics
is $N$ distinct notions of causal ordering.  We resolve this issue,
introduce coupling, and bring the model very close to already
well-established literature by demanding that each vertex's metric be
conformally related to its unique graph parent.  Let $p$ have a set of
children $\{a_1, \dots, a_{n_a}\}$.  Then, for the $i$-th child, there
exists a scalar field $\phi_{a_i}$ and constants $\sigma_{a_i}$ such that
\begin{align}
\xo{a_i}{g}_{\mu\nu} \equiv \exp(\sigma_{a_i}\phi_{a_i})\xo{p}{g}_{\mu\nu}
\label{eqn:conformal_relations}
\end{align}
where we have used the standard exponential parameterisation with the
factor of two removed for economy of notation.  Note that, from any
child's perspective, the parent metric acquires an inverted sign
\begin{align}
\xo{p}{g}_{\mu\nu} = \exp(-\sigma_{a_i}\phi_{a_i})\xo{a_i}{g}_{\mu\nu}.
\label{eqn:sign_inverse}
\end{align}
These $N-1$ scalar fields define the edges of the graph and replace
$N-1$ metric degrees of freedom.  

Note carefully that the model is no longer permutation symmetric.  In
other words, for distinct $p$, the resultant models from application
of the extrema principle will yield non-equivalent equations of
motion.  Nevertheless, the \emph{qualitative} behaviour observed in
all such models will be unchanged.  This can be seen by considering a
model where each vertex has a fixed number of children $n_a$ and the
tree is countably infinite.  In this limit, one would have perfect
permutation symmetry.  For a discussion of an absolutely permutation
invariant construction, and why we have avoided this model, we refer
the interested reader to Appendix~\S\ref{sec:foil_model}.  In the
remainder of our discussion, we will focus on a specific model
anchored at $p$.  We will use the word ``native'' to refer to tensors
associated with vertex $p$, and the word ``foreign'' to refer to
tensors associated with other vertices.  Special attention must be
paid to the scalar fields, however, as they define the graph's edges,
and are a property defined between two vertices.

\subsection{Asexual reproduction and scalar potentials}
\label{sec:sex}
Our next task is to implement the Smolin-like branching process, which
he assumes to be associated with black hole (BH) production.  The
approach we will take is distinctly pragmatic, and motivated from
condensed matter systems.  Consider the bulk magnetisation of a
ferromagnetic material as it passes through the Curie Temperature from
above.  Here, a vector degree of freedom enters the dynamics at a
phase transition.  In complete analogy, we assume that a scalar degree
of freedom $\phi_a$ and a tensor degree of freedom $T^a_{\mu\nu}$
enter the dynamics during gravitational collapse.  In this way, we
seek to implement a proposal of Chapline and Laughlin who, a decade
earlier, argued\cite{chapline2003quantum, chapline2001quantum} that
one really should replace the singular collapse scenario with a phase
transition.  We will glue an $N$ vertex model to a separate $N+1$
vertex model, both anchored at $p$, using the final data of the
$N$-vertex model fields to ``bootstrap'' the initial data of the $N+1$
model in a consistent fashion.

Chapline and Laughlin assert that the simplest scenario resulting from
their phase transition should be an interior GR vacuum solution with
anomalously large cosmological constant.  Adapted to our context, in
its simplest form, this can be accomplished by a single scalar field
dominated by potential.  Yet unlike the matter actions, which are
constrained by the Equivalence Principle; and the gravitational
actions, which are constrained by Boulanger, \emph{et. al.}; in the
absence of a comprehensive theory, a potential is explicitly
phenomenological.  There is an extremely rich literature on
inflationary potentials, their observational signatures, and their use
in resolving outstanding problems in particle physics (for an
excellent review, see \cite{bassett2006inflation}).  Inflationary
theory, however, is dominated by particle and quantum-field theoretic
concepts, and any foray would be premature given the present limited
scope.  Thus, we only partially constrain the specific form of $V$,
and instead focus on how potentials should enter our actions.

Using Rosen's notation\cite{rosen1995discrete}, let $[A_{qw}]$ be the
adjacency matrix for the ancestral tree.  Let $d(p,q)$ be the \emph{signed}
relative graph depth (c.f. $r(p,q)$) between vertex $p$ and vertex
$q$, with ancestors negative consistent with (\ref{eqn:sign_inverse}).
We define
\begin{align}
\xo{p}{S}_V \equiv - \frac{1}{8 \pi G}\int\rmd^4\xi~ \sum_{q}^N  \sum_{w > q} \kappa^{r(p,w)}[A_{qw}] V(\phi_w)\sqrt{w}
\label{eqn:pot_actions}
\end{align}
where the notation $w > q$ means to consider pairings such that $w$
satisfies $d(p,w) > d(p,q)$.  Note that $V: \mathcal{F}(\mathscr{M})
\leftarrow \mathcal{F}(\mathscr{M})$ is a fixed map understood to have
dimension of inverse length squared.  A single map is consistent with
the Copernican Principle and our restriction to a population of SM
clones.  In reheating scenarios, energy within the scalar field is
transferred to SM degrees of freedom.  The Equivalence Principle then
suggests an absence of cross-couplings between scalar fields in $V$.
Note again that, assuming no active scalar degrees of freedom, one
just regenerates $N$ distinct cosmological constants.

\subsection{A possible observational signature}
An interesting and immediate consequence of
(\ref{eqn:conformal_relations}) is that all members of the ancestral
tree share \emph{the same} (tensor) gravitational radiation field in
vacuum.  This is because the conformally invariant Weyl tensor is the
only non-zero component of the Riemann tensor in vacuum.  This is
consistent with the presence of a single metric degree of freedom,
whose quantized linearization would supposedly provide the graviton.
Since the radiation field is shared, a significantly higher than
expected observation rate for BH mergers
(e.g. \cite{abbott2016gw151226}) could be the result of foreign
sources.  Given sufficient directionality, at distances near enough to
the Earth for appreciable amplitude, a slightly aspherical supernova
gravitational wave signature coupled with a lack of observation in the
EM and neutrino sectors would be a ``smoking gun'' independent of
gravitational wave polarisation.

\section{The nearest-neighbour approximation}
\label{sec:nearest_neighbor_model}
To begin exploration of the phylogenic model, we work to first order
in $\kappa$, as this simplifies the resultant equations of motion and
facilitates preliminary confrontation with experiment.  To bring the
gravitational Lagrangian into standard multiscalar-tensor form, it is
most natural to work with $p$'s metric.  Let $q$ index over $N-1$
vertices adjacent to $p$.  We define $x$ to be $p$'s parent, and $a$
to be a numeric index over $p$'s children, where we drop numeric
subscripts because no ambiguity can arise in this limited context.  By
the conformal relations (\ref{eqn:conformal_relations}), we have
that
\begin{align}
\xo{p}{g}_{\mu\nu} &\equiv \exp(\sigma_p\phi_p) \xo{x}{g}_{\mu\nu}  &\xo{a}{g}_{\mu\nu} &\equiv \exp(\sigma_a\phi_a)\xo{p}{g}_{\mu\nu} \\
\sqrt{p} &= \exp(2\sigma_p\phi_p)\sqrt{x}  & \sqrt{a} &= \exp(2\sigma_a\phi_a)\sqrt{p}
\end{align}
which is visualised in Figure~\ref{fig:tree_schematic}.
Recall\cite{carroll2004spacetime} that Ricci scalars for conformally
related metrics in $n$ spacetime dimensions satisfy
\begin{align}
\xo{a}{R} = \exp(-\sigma_a\phi_a)\left[\xo{p}{R} + (n-1)\left(\frac{1}{4}\sigma_a^2(2-n)\xo{p}{\nabla^\nu}\phi_a\xo{p}{\nabla_\nu}\phi_a - \sigma_a\xo{p}{\nabla^\nu}\xo{p}{\nabla_\nu} \phi_a\right)\right].
\label{eqn:conformal_ricci}
\end{align}
We then find terms of the form
\begin{align}
\xo{a}{R}\sqrt{a} = \exp(\sigma_a\phi_a)\left[\xo{p}{R} + (n-1)\left(\frac{1}{4}\sigma_a^2(2-n)\xo{p}{\nabla^\nu}\phi_a\xo{p}{\nabla_\nu}\phi_a - \sigma_a\xo{p}{\nabla^\nu}\xo{p}{\nabla_\nu} \phi_a\right)\right]\sqrt{p}
\end{align}
appearing in the gravitational action.  We may now integrate by parts, exploiting
\begin{align}
\exp(\sigma_a\phi_a)\xo{p}{\nabla^\nu}\xo{p}{\nabla_\nu}\phi_a  = \xo{p}{\nabla_\mu}\left[\xo{p}{g^{\mu\nu}}\exp(\sigma_a\phi_a)\xo{p}{\nabla_\nu}\phi_a\right] - \sigma_a\xo{p}{\nabla^\nu}\phi_a\xo{p}{\nabla_\nu}\phi_a\exp(\sigma_a\phi_a)
\end{align}
to remove the second derivatives.  This results in integrand terms of the form
\begin{align}
\xo{a}{R}\sqrt{a} = \exp(\sigma_a\phi_a)\left[\xo{p}{R} + \frac{\sigma_a^2}{4}(n-1)(n-6)\xo{p}{\nabla^\nu}\phi_a\xo{p}{\nabla_\nu}\phi_a\right]\sqrt{p} + \partial\sqrt{p}
\end{align}
where $\partial$ is a total divergence.  Substitution into
(\ref{eqn:grav_actions}), dropping $\partial$, gives the following
gravitational Lagrangian density
\begin{align}
\xo{p(1)}{\mathscr{L}_G} = &\left\{\xo{p}{R}\left[1 + \kappa\sum_q\xo{q}{\exp}(\phi\sigma)\right] - \sum_{q,w} \nabla^\nu\xo{q}{\phi}\nabla_\nu\xo{w}{\phi}\left[\frac{\kappa}{4}(n-1)(n-6)\xo{q}{\sigma^2} \xo{q}{\exp}(\phi\sigma)\delta^{qw}\right]\right\}\sqrt{p} + O(\kappa^2).
\label{eqn:grav_lagrangian_jordan}
\end{align}
We have omitted labels over the covariant derivatives, since they are
always with respect to $p$, and overset the exponential to indicate
that its arguments are regarded as particular to that specific label.
Note that (\ref{eqn:grav_lagrangian_jordan}) is expressed in the
Jordan frame.  Consistent with the notation of
\cite{kuusk2016invariant} we define
\begin{align}
A(\phi_q, \dots) &\equiv 1 + \kappa\sum_q\xo{q}{\exp}(\phi\sigma) + O(\kappa^2)\label{eqn:A}\\ 
B^{qw}(\phi_q, \dots) &\equiv \frac{\kappa}{4}(n-1)(n-6)\xo{q}{\sigma^2} \xo{q}{\exp}(\phi\sigma)\delta^{qw} + O(\kappa^2)\label{eqn:B}.
\end{align}
Note that terms of order $\kappa^2$ and higher would produce a
non-diagonal matrix in (\ref{eqn:B}).  This is because more distant
``relatives'' would be conformally related to $p$ by terms
$\exp(\sum_q \sigma_q\phi_q)$.  Upon differentiation, these would introduce
mixed products of covariant derivatives depending on the specific
relative.

\subsection{Classical instabilities}
A common pathology in classical field theories is unbounded energy
exchange, also called classical instability.  Independently of the
matter sector, a multiscalar-tensor theory is
unstable\cite{kuusk2016invariant} if
\begin{align}
\frac{B^{qw}}{2A} + \frac{3}{4A^2}\frac{\partial A}{\partial \phi_q}\frac{\partial A}{\partial \phi_w} \geq 0.
\end{align}
Note that the above expression becomes the coefficient on the kinetic
terms when the multiscalar-tensor theory is cast into the Einstein
frame.  Note that $A$ is always non-zero if
\begin{align}
\kappa \ll 1/N
\label{eqn:kappa_constraint}
\end{align}
and that this is equivalent to saying that there are not ``too many''
offspring for the given $\kappa$.  In this scenario, we may multiply
through by $2A$
\begin{align}
B^{qw} + \frac{3}{2A}\frac{\partial A}{\partial \phi_q}\frac{\partial A}{\partial \phi_w} \geq 0.
\label{eqn:ghostbuster}
\end{align}
At order $\kappa$, only the $B^{qw}$ term contributes provided that
the second term remains bounded.  Consider the diagonal where $q=w$.
Immediately, we must have $\forall q$ that
\begin{align}
\sigma_q \equiv 
\left\{ 
\begin{array}{cc}
iy_q & 1 < n < 6 \\
y_q & n \geq 7 
\end{array}\right.\qquad y_q \in \mathbb{R}
\end{align}
and we have arrived at a rather curious conclusion.  For spacetime
dimension $n=4$, each metric tensor in the population must be related
to every other by a pure phase.  Paying attention to the real portion
of (\ref{eqn:ghostbuster}), we also require that
\begin{equation}
\phi_q \in \left[-\pi/2, \pi/2\right] + 2\pi m \qquad m \in \mathbb{Z}
\label{eqn:phi_constraint}
\end{equation}
be dynamically enforced.  Investigation of the second term in
(\ref{eqn:ghostbuster}) gives
\begin{align}
-\frac{3\kappa^2\xo{q}{\exp}(2\sigma\phi)\left[1 + \kappa\sum_n\xo{n}{\exp}(-2iy\phi)\right]}{2 + 2\kappa^2 N + 4\kappa^2\sum_{n\neq m}\cos(y_n\phi_n - y_m\phi_m)}
\end{align}
where we have made the denominator real.  This term remains
$O(\kappa^2)$ provided that $\kappa \ll 1/\sqrt{N}$, which is
consistent with (\ref{eqn:kappa_constraint}).  The off-diagonal terms
remain similarly bounded, and can thus be discarded in the present
treatment.  We conclude the nearest-neighbour model is free of
instabilities in spacetime dimension $n=4$, provided that metrics are
conformally related by a pure phase and (\ref{eqn:phi_constraint})
remains satisfied.  The presence of this phase leads to many
stimulating questions of interpretation.  Unfortunately, these
questions would lead us too far from our present scope.  For
simplicity, we now focus on the model got by demanding the real
variation of the action to vanish identically, while permitting the
imaginary portion to float according to the solutions of the real
system.  For discussion of the $n\geq 7$ phylogenic model, which does
not require the introduction of any complex numbers, we refer the
interested reader to Appendix~\S\ref{sec:alternative}.

We must now define our approach to invariant spacetime lengths.  We
proceed by considering the effect of conformally scaling the usual
spatially flat Robertson-Walker (RW) metric
\begin{align}
\xo{p}{\rmd s^2} = -\rmd t^2 + p(t)^2\rmd \vec{x}^2
\end{align}
where $p(t)$ is vertex $p$'s scale factor, as reckoned from observers
within $p$.  Now consider a single child labelled $q$, and choose the
convention that $\sigma_q \equiv +i$.  By
(\ref{eqn:conformal_relations}), we have that
\begin{align}
\xo{q}{\rmd s^2} = -\exp(i\phi_q)\rmd t^2 + \exp(i\phi_q)p(t)^2\rmd \vec{x}^2
\end{align}
the real portion of which is
\begin{align}
\xo{q}{\rmd s^2} = -\cos(\phi_q)\rmd t^2 + \cos(\phi_q)p(t)^2\rmd \vec{x}^2.
\end{align}
Observers in $q$ will proceed to make physical conclusions with \emph{their} most natural time coordinate
\begin{align}
\tau(t) \equiv \int_c^t \sqrt{\cos(\phi_q(t'))}~\rmd t' \implies \rmd \tau = \sqrt{\cos(\phi_q(t))}~\rmd t
\label{eqn:foreign_time}
\end{align}
Thus, observers in $q$ simply perceive the usual RW metric
\begin{align}
\xo{q}{\rmd s^2} \equiv -\rmd \tau^2 + q(\tau)^2\rmd \vec{x}^2 \label{eqn:foreign_rw}
\end{align}
while observers in $p$ ``know better'' 
\begin{align}
q(\tau) \equiv p\sqrt{\cos(\phi_q)}\label{eqn:foreign_scale}.
\end{align}
Note that our definition (\ref{eqn:foreign_time}) again requires that
(\ref{eqn:phi_constraint}) remain satisfied at all times.  We thus see
that an exchange of the spacelike and timelike trajectories signals
the emergence of a classical instability.

By the Copernican Principle encoded into the gravitational action
(\ref{eqn:grav_actions}), we expect that offspring should begin from a
very tiny spatial point since this is how we appear to have begun.
Given (\ref{eqn:foreign_rw}) and (\ref{eqn:foreign_scale}), this
occurs at some branching time $t_*$, unique to each offspring, if
\begin{align}
\phi_{q}(t_*) \equiv \frac{(2m + 1)\pi}{2} \qquad m\in\mathbb{Z}.
\label{eqn:position_ic}
\end{align}
Near these points, the child's clock is significantly slowed from the
parent's perspective.  For the $n=4$ model, this effect is actually
\emph{symmetric} along the ancestral tree between any two vertices,
since cosine is even.  This is reminiscent of clock rates measured by
observers in relative inertial motion under Special Relativity.

\subsection{Metric equations of motion}
\label{sec:eom}
Having established that the nearest-neighbour model is free of
instabilities if (\ref{eqn:phi_constraint}) remains satisfied, we
proceed to compute the equations of motion in $n=4$ spacetime
dimensions.  Consequently, in all equations below, $\sigma \equiv \pm
i$.  Sign is of course determined by ancestry and $y \equiv 1$ has
been assumed without loss of generality.  Note that care must be taken
to not prematurely divide out exponential factors, as their real
portions may evaluate to zero.  Equations of motion will be found from
requiring that the real portion of
\begin{align}
\delta S \equiv \delta \xo{p}{S}_G + \delta \xo{p}{S}_M + \delta \xo{p}{S}_V
\end{align}
vanish.  Substitution of ancestry (\ref{eqn:conformal_relations}) into
the matter actions (\ref{eqn:matter_actions}) through order $\kappa$
gives
\begin{align}
\delta S_M = -\frac{1}{2}\int \rmd^4\xi~\left\{\left[\xo{p}{T}_{\mu\nu} + \kappa\sum_q\xo{q}{\exp}(\sigma\phi) \xo{q}{T}_{\mu\nu}\right]\delta\xo{p}{g^{\mu\nu}} - \kappa\sum_q\xo{q}{\sigma}\xo{q}{\exp}(2\sigma\phi)\xo{q}{T}\delta\xo{q}{\phi}\right\}\sqrt{p} + O(\kappa^2).
\end{align}
To determine the potential contribution, we first expand the adjacency
sum of (\ref{eqn:pot_actions}) to first order in $\kappa$
\begin{align}
\xo{p}{S}_V = -\frac{1}{8\pi G}\int \rmd^4\xi~\left[ \kappa \xo{p}{V}\sqrt{p} + \kappa\sum_a \xo{q}{V}\sqrt{a}\right] + O(\kappa^2),
\end{align}
substitute ancestry, and perform the variation
\begin{align}
\delta S_V = \frac{\kappa}{16\pi G}\int \rmd^4\xi~\left\{\delta\xo{p}{g^{\mu\nu}}\left[\xo{p}{V}\xo{p}{g_{\mu\nu}} + \xo{p}{g_{\mu\nu}}\sum_q\xo{q}{\exp}(2\sigma\phi)\xo{q}{V}\right] - 2\frac{\partial \xo{p}{V}}{\partial \phi}\delta\xo{p}{\phi} - 2\sum_q\xo{q}{\exp}(2\sigma\phi)\left[\frac{\partial \xo{q}{V}}{\partial \phi} + 2\sigma \xo{q}{V}\right]\delta\xo{q}{\phi}\right\}\sqrt{p}.
\end{align}

We leverage the results of \cite{kuusk2016invariant} to write down the
metric and scalar equations of motion using (\ref{eqn:A}) and
(\ref{eqn:B}).  We find for the mixed metric equation of motion
\begin{align}
\left(1+\kappa \sum_q \xo{q}{\exp}(\sigma\phi)\right)G^\mu_\nu - \kappa\sum_q \xo{q}{\exp}(\sigma\phi) \left[\left(\frac{\nabla^\alpha\xo{q}{\phi} \nabla_\alpha\xo{q}{\phi}}{4} - \xo{q}{\sigma} \nabla^\alpha\nabla_\alpha \xo{q}{\phi} \right)\delta^\mu_\nu  + \frac{\nabla^\mu\xo{q}{\phi}\nabla_\nu\xo{q}{\phi}}{2}  + \xo{q}{\sigma}\nabla^\mu\nabla_\nu\xo{q}{\phi} \right]& \nonumber \\
 - 8\pi G\xo{p}{T^\mu_\nu} +\kappa \xo{p}{V}\delta^\mu_\nu +\kappa\sum_q\xo{q}{\exp}(2\sigma\phi)\left[\xo{q}{V}\delta^\mu_\nu  - 8\pi G\xo{q}{T^\mu_\nu}\right] + O(\kappa^2)& = 0
\label{eqn:metric_eom}
\end{align}
where we are careful to raise the index on the foreign stress after we
have changed to the foreign metric, which introduces a conformal
factor.  At this point, recall the considerations of \S\ref{sec:sex},
where we note that the dynamics must not become acausal when a scalar
field enters during the collapse process.  In order that the effective
gravitational constant remain initially unchanged, we have immediately
that
\begin{align}
\xo{q}{\phi}(\vec{x}, t_*) \equiv \frac{\pi}{2} + 2\pi m \qquad m \in \mathbb{Z}
\label{eqn:birth_ic}
\end{align}
which is nothing more than the generalisation of
(\ref{eqn:position_ic}) to arbitrary metrics.  Since
(\ref{eqn:kappa_constraint}) guarantees that we do not have excessive
offspring, it is useful to divide the metric equations
(\ref{eqn:metric_eom}) by $A(\phi, \dots)$ and Taylor expand through
$O(\kappa)$
\begin{align}
G^\mu_\nu - \kappa\sum_q \xo{q}{\exp}(\sigma\phi) \left[\left(\frac{\nabla^\alpha\xo{q}{\phi} \nabla_\alpha\xo{q}{\phi}}{4} - \xo{q}{\sigma} \nabla^\alpha\nabla_\alpha \xo{q}{\phi} - \xo{q}{\exp}(\sigma\phi)\xo{q}{V}\right)\delta^\mu_\nu  + \frac{\nabla^\mu\xo{q}{\phi}\nabla_\nu\xo{q}{\phi}}{2}  + \xo{q}{\sigma}\nabla^\mu\nabla_\nu\xo{q}{\phi}\right]& \nonumber \\
 - 8\pi G\xo{p}{T^\mu_\nu} + \kappa \xo{p}{V}\delta^\mu_\nu- 8\pi G\kappa\sum_q\xo{q}{\exp}(\sigma\phi)\left[\xo{q}{\exp}(\sigma\phi)\xo{q}{T^\mu_\nu} - \xo{p}{T^\mu_\nu}\right] + O(\kappa^2)& = 0.
\label{eqn:metric_eom_einstein}
\end{align}
This equation is Einstein's equation with a position dependent
gravitational constant (somewhat obscured by grouping), augmented with
scalar kinetic and potential terms.  A new feature due to our matter
actions is the presence of foreign stress, which appears here as
purely dark matter.

\subsection{Scalar equations of motion and screening}
\label{sec:mimetic}
Suppressing $q$ labels on $\phi$ for clarity, since there is no
cross-coupling between the scalar fields at order $\kappa$, the scalar
equation of motion for each $q$ is found to be
\begin{align}
\kappa\xo{q}{\exp}(\sigma\phi)&\left(\xo{q}{\sigma}\xo{p}{R} + 3\nabla^\alpha\nabla_\alpha\phi + \frac{3\xo{q}{\sigma}}{2}\nabla^\alpha\phi\nabla_\alpha\phi \right) + 8\pi G\xo{q}{\sigma}\kappa\xo{q}{\exp}(2\sigma\phi)\xo{q}{T} \nonumber \\
& - 2\kappa\left[\frac{\xo{q}{\partial V}}{\partial \phi}\delta^p_q + (1-\delta^p_q)\xo{q}{\exp}(2\sigma\phi)\left(2\sigma \xo{q}{V} + \frac{\xo{q}{\partial V}}{\partial \phi}\right) \right] + O(\kappa^2) = 0.
\label{eqn:scalar_eom_ricci_child}
\end{align}
We now take the trace of the full metric EOM
(\ref{eqn:metric_eom})
\begin{align}
-\xo{p}{R}\left(1 + \kappa\sum_q\xo{q}{\exp}(\sigma\phi)\right) + &3\kappa\sum_q\xo{q}{\exp}(\sigma\phi)\left[\xo{q}{\sigma}\nabla^a\nabla_\alpha\xo{q}{\phi} - \frac{\nabla^\alpha\xo{q}{\phi}\nabla_\alpha\xo{q}{\phi}}{2}\right] \nonumber \\
&- 8\pi G\xo{p}{T} + 4\kappa\xo{p}{V} + \kappa\sum_q\xo{q}{\exp}(2\sigma\phi)\left[4V(\xo{q}{\phi}) - 8\pi G\xo{q}{T}\right] + O(\kappa^2) = 0.
\label{eqn:eom_trace}
\end{align}
and use it to remove the Ricci scalar to leading order
\begin{align}
\xo{q}{\exp}(\sigma\phi)\left[3\nabla^\alpha\nabla_\alpha\phi + \frac{3\sigma}{2}\nabla^\alpha\phi\nabla_\alpha\phi\right] &= \nonumber \\
8\pi G\xo{q}{\sigma}\xo{q}{\exp}(\sigma\phi)&\left[\xo{p}{T} - \xo{q}{\exp}(\sigma\phi)\xo{q}{T}\right] + 2\left[\frac{\xo{q}{\partial V}}{\partial \phi}\delta^p_q + (1-\delta^p_q) \xo{q}{\exp}(2\sigma\phi)\left(2\xo{q}{\sigma} \xo{q}{V} + \frac{\xo{q}{\partial V}}{\partial \phi}  \right)\right]+ O(\kappa).
\label{eqn:scalar_eom}
\end{align}
Note that the potential contribution distinguishes the scalar
equations, depending on relation within the ancestral tree.  In the
absence of any potential contribution, (\ref{eqn:scalar_eom}) takes
the familiar form of a curved-space wave equation with kinetic drag,
with the marked exception that the source term can \emph{zero-cross}.
Sign alteration of the source about
\begin{align}
\exp(\sigma \phi)\xo{p}{T} = \exp(2\sigma \phi)\xo{q}{T}
\end{align}
will drive the field to act as if it were free.  We are unaware of any
form of screening in the existing literature with this behaviour.
Note that screening is only possible for $\phi \in [0, \pi/4]$ if both
sources respect (or violate) the strong energy condition and
\begin{align}
|\xo{p}{T}| \leq |\xo{q}{T}|.
\end{align}
If $\phi \in [\pi/2, \pi/4)$ one, but not both, of the sources must
  violate the strong energy condition in order screen.  We digress
  momentarily from the $n=4$ theory to emphasise that screening
  persists in the purely real $n \geq 7$ theory.

\subsection{Branching consistency and initial conditions}
Since the equations of motion (\ref{eqn:metric_eom}) obtain from a
scalar action, they are automatically covariantly conserved.  This is
true for both an $N$ vertex model anchored at $p$ and an $N+1$ vertex
model anchored at $p$.  Since transitioning between these two models
at a collapse event must not violate local conservation, we require
that the covariant divergence of the newly introduced terms must
vanish.  Assign the label $q$ to the newly introduced vertex.  Then
consistency requires that
\begin{align}
\sigma\nabla_\mu\phi\Lambda^\mu_\nu + \nabla_\mu\Lambda^\mu_\nu =\Bigg|_{t_*} \exp(\phi \sigma)\left[8\pi G \xo{p}{\nabla}_\mu\xo{q}{T^\mu_\nu} - \nabla_\nu V + 2\sigma\nabla_\mu\phi\left(8\pi G\xo{q}{T^\mu_\nu} - V\delta^\mu_\nu\right)\right]
\label{eqn:cov_birth_n4}
\end{align}
where $\Lambda^\mu_\nu$ is defined in (\ref{eqn:Lambda}) and the $q$
label has been suppressed except in locations of possible confusion.
By inspection, (\ref{eqn:cov_birth_n4}) simplifies
considerably if the scalar field is initially gradient-free
\begin{align}
\left.\nabla_\mu \xo{q}{\phi}(x^\alpha)\right|_{t_*}  \equiv 0.
\label{eqn:velocity_ic}
\end{align}
Note that this initial condition leaves the initial value of the
potential and its derivative unconstrained (so long as they are
finite).  Substitution of (\ref{eqn:velocity_ic}) into
(\ref{eqn:cov_birth_n4}) gives
\begin{align}
\nabla_\mu\Lambda^\mu_\nu =\Bigg|_{t_*} 8\pi G \xo{p}{\nabla}_\mu\xo{q}{T^\mu_\nu}\exp(\phi \sigma)
\end{align}
where the gradient of $V$ has vanished because $V$ produces
real-valued $\mathcal{F}(\mathscr{M})$, whereas re-expression of the
covariant derivative will introduce metric factors.  Focusing on the
real portion removes the covariant accelerations, and using the
contracted Bianchi identity we find
\begin{align}
\frac{1}{4}\nabla_\nu\left[\nabla^\alpha\phi\nabla_\alpha\phi\right] + \frac{1}{2}\nabla_\mu\left[\nabla^\mu\phi\nabla_\nu\phi\right] =\Bigg|_{t_*} 4\pi G \left(\xo{q}{T^\alpha_\nu}\nabla_\alpha\phi  + \xo{q}{T}\nabla_\nu\phi\right)\sin(\phi).
\end{align}
Since each term on the left will contain a gradient of the field after
differentiation, we see covariant conservation is satisfied during
branching given the initial condition (\ref{eqn:velocity_ic}).  That
this initial condition is sufficient should not be surprising: energy
transfer associated with a collapse event should be (initially)
spatiotemporally localised to the collapse.

If we evaluate the parent scalar equation of motion at the initial
condition (\ref{eqn:birth_ic}), we find
\begin{align}
4\pi G\xo{p}{T} =\Bigg|_{t_*} \frac{\partial \xo{p}{V}}{\partial \phi}.
\end{align}
For all known matter and radiation, the left hand side is always less
than or equal to zero.  The analogous relation for an offspring scalar
equation at the initial condition introduces a sign.  Since $V$ is a
fixed map, this means that $V$ must have an extremum at the initial
condition.  Note that during the usual cold \emph{and} warm inflation
scenarios, the left hand side vanishes, which is consistent with the
extremum.  Further, this corresponds to a constant potential in the
neighbourhood of the initial condition, which suggests exponential
inflation.

\section{Newtonian behaviour}
\label{sec:consistency}
In this section, we begin the first investigations of whether the
nearest-neighbour model can reproduce the phenomenology of GR on solar
system scales.  First we will compute the Newtonian limit, and
highlight the promising screening behaviour.  Unfortunately, at this
point, we cannot use the well-known results of Esposito-Farese and
Damour\cite{damour1992tensor} to draw PPN conclusions because we have
multiple stress-energies.

We approximate solutions to scenarios featuring weak-fields
\begin{align}
V(\phi) \sim \epsilon ^2 &\qquad \frac{\partial V}{\partial \phi} \sim \epsilon ^2 \\
\xo{p}{\rho}(x^\alpha) \sim \epsilon ^2 &\qquad \xo{q}{\rho}(x^\alpha) \sim \epsilon^2\\ 
\xo{p}{g}_{\mu\nu} = \eta_{\mu\nu} + \epsilon^2 \xo{p}{h_{\mu\nu}}(x^\alpha) + O(\epsilon^4) &\qquad \xo{q}{\phi}(x^\alpha) = \xo{q}{\Phi} + \epsilon^2 \xo{q}{\phi}(x^\alpha) + O(\epsilon^4).
\end{align}
This is most appropriate for a parent $p$ considering contributions
from offspring $q$.  The Newtonian regime is additionally
characterised by slow motions
\begin{align}
\frac{\partial}{\partial t} \sim \epsilon \frac{\partial}{\partial x}.
\end{align}
We will regard the scalar contributions as a source, and must be
careful to consider their entire active gravitational mass
i.e. effective mass density plus spatial trace.  Using Will's
notation\cite{TEGP} for the gravitational potential, in this
approximation, combination of (\ref{eqn:metric_eom}) and
(\ref{eqn:scalar_eom}) gives the following Poisson equation after
taking real portions
\begin{align}
\nabla^2 U = 4\pi G \left[\xo{p}{\rho} - \sum_q\kappa\cos(\xo{q}{\Phi})\left(\xo{p}{\rho} - \xo{p}{T}/6\right) \right] + \kappa\sum_q\left\{4\pi G\cos(2\xo{q}{\Phi})\left[\xo{q}{\rho} - \xo{q}{T}/6\right] + \frac{1}{6}\left[\frac{\partial V}{\partial \phi}\sin(2\xo{q}{\Phi}) - \cos(2\xo{q}{\Phi})V\right]\right\}
\label{eqn:newtonian_limit}
\end{align}
Neglecting native pressures, this is Newton's equation with a weakened
gravitational constant, dark matter, and scalar potential.  We now
specialise this result to some relevant limits, and focus only on the
behaviour of a single contributing offspring $q$.

\subsection{Contribution to $p$ from inflating offspring}
We will assume the cold inflationary scenario for our single offspring
$q$.  Of course, if $q$ inflates, we expect $V$ to be of the same
order as the collapsed parent matter, so $V \sim O(1)$.  To get a feel
for what happens in this limit at a distance from the collapsed
object, we formally adjust $\kappa$ to guarantee the validity of the
Newtonian approximation from $p$'s perspective.  Removing all stress
from (\ref{eqn:newtonian_limit}), we find
\begin{align}
\nabla^2 U = \frac{\kappa}{6}\left[\frac{\partial V}{\partial \phi}\sin(2\xo{q}{\Phi}) - \cos(2\xo{q}{\Phi})V\right].
\end{align}
Since at $t_*$ the offspring is effectively
frozen from the parent's perspective, we find that
\begin{align}
\nabla^2 U = \frac{\kappa}{6} V\left(\frac{\pi}{2}\right).
\end{align}
 We emphasise strongly that we have not yet investigated the radial
 behaviour of the nearest-neighbour model in the static, spherically
 symmetric limit.  This limit should be the most relevant after
 collapse due to the relative clock rates between parent and
 offspring, and is required to confirm the expected localised
 quasi-static behaviour of a Chapline-Laughlin BH.

\subsection{Contribution to $p$ from late-stage offspring}
\label{sec:screening}
After the offspring has reheated, we expect the potential to no longer
contribute substantially to the equations of motion.  Since we
(present epoch observers) are stationed in $p$, it is reasonable that
$p$'s pressure be zero.  Due to the difference in clock rates, we
might expect vertex $q$ to be radiation dominated.  If we set $V$ and
$q$'s trace to zero in (\ref{eqn:newtonian_limit}) we find
\begin{align}
\nabla^2 U = 4\pi G \xo{p}{\rho}\left[1 - \frac{7\kappa}{6}\cos(\xo{q}{\Phi})\right] + \kappa4\pi G\cos(2\xo{q}{\Phi})\xo{q}{\rho}.
\label{eqn:newtonian_after_inflation}
\end{align}
Note that since we have assumed $\Phi \in [0, \pi/2]$, the dark matter
term can enter with \emph{repulsive} character if $\Phi > \pi/4$.
Such repulsive matter contributions on large scales are disfavoured by
the lack of deviations between positive mass simulations and
observation in the 3rd moment (and thus sensitive to sign) of the mass
convergence maps from weak-lensing surveys\cite{heymans2012cfhtlens}.
Yet, any means to realise ``negative energy'' without instability is
interesting.  For $\Phi \in [0, \pi/4]$, the scalar field will
oscillate about
\begin{align}
\frac{\cos(\Phi)}{\cos(2\Phi)} = \xo{q}{T}/\xo{p}{T}
\label{eqn:mimetic_condition}
\end{align}
given the constraints discussed in \S\ref{sec:mimetic}.  Since the
source term is essentially a driving force to a harmonic oscillator
with kinetic drag, this suggests damped oscillation about an
equilibrium point within this range.  More so, the equilibrium point
will track the evolving matter densities.  Remarkably, at exactly this
equilibrium point (\ref{eqn:newtonian_limit}) becomes
\begin{align}
\nabla^2 U = 4\pi G \xo{p}{\rho},
\end{align}
which is precisely Newton's equation.  

\section{Discussion and future directions}
We have chosen to investigate the $n=4$ theory primarily because we
find ourselves within a 3+1 dimensional spacetime.  For the purpose of
model building, however, the pure phase model is awkward due to the
introduction of complex quantities.  Yet the pure phase model is
demonstrably perturbative in $\kappa$, and this permits calculation of
the condition under which the nearest-neighbour approximation remains
dynamically stable.  Whether this dynamical condition is honoured
provides the first target for subsequent cosmological investigations.

While it may be plausible that the stability condition be honoured by
a single contributing field, the superposition of many fields in the
$n=4$ theory, as would be required at higher order in $\kappa$, is
undesirable.  Since dynamical stability in the $n=4$ model is linked
to the preservation of spacetime interval sign, the
positive-semidefinite conformal factors required in models with $n\geq
7$ motivate a complete stability analysis of the full phylogeny.

An attractive feature for all models, independent of $n$, is that
consistency of the branching process requires that offspring universes
begin at a spatial point.  This ``big bang'' condition, though highly
dynamical from the perspective of the offspring, may appear initially
static from the perspective of the parent.  The behaviour of static,
spherically symmetric, solutions in the $N=2$ case from the parent's
perspective provides an attractive target for future investigations.

PPN constraint is a more demanding calculation.  Present frameworks
for PPN of multiscalar-tensor theories with potential are highly
non-trivial\cite{HohmannErik}, as is the PPN of theories with multiple
sources of stress (e.g. \cite{HohmannPPN}).  Given the highly
constrained nature of the model presented, encouraging cosmological
and spherically symmetric results should precede any PPN
investigation.  A reasonable first step would depart from the
brute-force perturbation pioneered by Will and Nordvedt\cite{TEGP}
toward a generalisation of Esposito-Farese's PPN technique to multiple
matter sources.  Inclusion of a potential would only be warranted if
cosmological and spherical investigations gave mandate.

It is very interesting that the model naturally features both
collisional and collisionless dark matter.  Collisional dark matter
(e.g. \cite{mahdavi2007dark}) would be inferred from dynamics of
source distributions within a single foreign contribution.
Collisionless dark matter (e.g. \cite{clowe2006direct}) would be
inferred from the dynamics of source distributions from multiple
foreign contributions.  These sources will not, in general, follow
geodesics of the native metric.  They will, however, alter native null
geodesics (i.e. lens) in the usual way.  The model also naturally
``tracks'' the native source.  Though at order $\kappa$ this tracking
is exact and hides the foreign source, at higher order this may no
longer be the case.  It should be noted that such ``tracking of the
luminous mass'' was an oft cited virtue of MOND-like phenomenologies
(e.g. \cite{milgrom1983modification, bekenstein2004relativistic}).
Cosmological constraints on Dark Matter, however, are severe; the
ratio of baryonic density to non-baryonic density cannot change.  This
would seem to exclude dominant parent and offspring contributions to
the Dark Matter density.  Sibling contributions at order $\kappa^2$,
however, provide an attractive candidate.  This is because the
Copernican Principle suggests that sibling growth and development
closely track our own.  Clearly, further exploration is justified.

\section{Conclusions}
We have implemented a Smolin-like branching multiverse as a
multiscalar extension to GR.  Our implementation seeks to produce a
more comprehensive model and remedy the lack of both population and
interaction within Smolin's original proposal.  We take the minimum
viable $N$ metric theory, in which metric degrees of freedom are
entirely decoupled, and replace $N-1$ metrics with scalar fields via
conformal relations.  The conformal relations enforce a directed,
acyclic graph structure upon the population of universes, i.e. a tree.
The result is a classical multiscalar-tensor field theory, and thus
amenable to experimental confrontation.  We analyse the model in $n$
spacetime dimensions and focus on a nearest-neighbour approximation
with $n=4$.  We determine the conditions for dynamical stability of
the this model and compute equations of motion.  The scalar equation
of motion exhibits a novel screening property: the field actively
seeks to decouple from stress under certain conditions.  We detail how
to consistently transition between an $N-1$ scalar field model to an
$N$ scalar field model, as would be required to guarantee well-defined
dynamics during reproduction events.  We compute the Newtonian limit
and show that, when applicable, the screening property reproduces
exactly Newton's equations.

\begin{acknowledgments}
The author would like to warmly thank Joel Weiner for guidance,
encouragement, and the crucial suggestion of conformal relation to
preserve causality.  The author further thanks Manuel Hohmann for
constructively critical discussions concerning early versions of this
work.  Many algebraic manipulations were verified using GNU Maxima.
Portions of this work were performed at the University of Tartu as a
Fulbright Fellow under the generous hospitality of the Laboratory of
Theoretical Physics, with financial support from the Fulbright
U.S. Student Program and the University of Hawai`i.
\end{acknowledgments}

\begin{appendix}
\section{Foil from absolute permutation symmetry}
\label{sec:foil_model}
In this appendix, we briefly discuss the following gravitational and
matter actions
\begin{align}
\delta \xo{p}{S}_M &\equiv -{1\over 2}\int\rmd^4\xi~ \sum_{q}^N  \kappa^{d(p,q)}\xo{q}{T}_{\mu\nu}\delta\xo{q}{g^{\mu\nu}}\sqrt{q} \\
\xo{p}{S}_G &\equiv \frac{1}{16\pi G}\int\rmd^4\xi~ \sum_{q}^N  \kappa^{d(p,q)} \xo{q}{R}\sqrt{q}
\end{align}
where signed relative graph depth \emph{replaces} the relative graph
distance $r(p,q)$.  Even with conformal constraint, the resultant
model is independent of the choice of $p$.  This can be seen by
noticing that a model anchored at $p$ and one anchored at any other
vertex are related by a scaling of $\kappa$ to some power.  This power
can be absorbed into the units of Newton's constant.

This absolute permutation symmetry might initially seem attractive.
We have avoided it for the following reasons.  Due to the symmetry,
the model can be analysed by anchoring within a leaf of the tree.
Then all couplings become $\kappa$ to some negative power, so let
$\kappa > 1$.  Unfortunately, all leaves enter at unit strength and
the notion of ``sibling'' is destroyed.  Similar relative couplings
across the tree grossly violate our intuitive notion of ``ancestry.''
For example, a newborn child in Australia should not influence a
similarly aged child born in Iceland more than either of their
parents, especially if this child were born two hundred years later.
Since the microevolutionary process works ``locally'' within ancestral
communities, the rigidly permutation symmetric model fails to capture
the essential aspects of actual biological populations.

\section{Conformal relation with the reals}
\label{sec:alternative}
In this appendix, we describe the essential differences between the
$n=4$ and $n \geq 7$ models.  We then briefly present the field
equations and the branching constraint in $n$ dimensions.

The essentially changed features are:
\begin{itemize}
\item{Spacetime dimension $\geq 7$: the actual dimension of spacetime
  must exceed 6.  This is not a constraint on the \emph{active}
  spacetime dimensions, and the usual compactification evasions may be
  employed.}
\item{Foreign time coordinates are always well-defined.}
\item{The partial order of the ancestral tree becomes explicitly
  physical.  An ancestor clock always runs faster, while a child clock
  always runs slower.  This can be used to define a conceptually
  distinct ``arrow of time.''}
\item{Perturbative treatment in $\kappa$ not guaranteed: the theory
  splits into two regimes, an Einstein regime similar to the pure
  phase model, and a ``nascent'' regime where the scalar field
  dominates the curvature.}
\item{Classical instabilities cannot be investigated perturbatively:
  since the perturbative expansion in $\kappa$ is no longer always
  valid, one must work to higher order in $\kappa$ to investigate
  whether terms remain positive.}
\end{itemize}
We proceed with little discussion.  The mixed metric equation of
motion is
\begin{align}
\left(1 + \kappa\sum_q\xo{q}{\exp}(\phi\sigma)\right)G^\mu_\nu + \kappa\sum_q\xo{q}{\sigma}^2\xo{q}{\exp}(\phi\sigma)\left[\delta^\mu_\nu\left(\frac{n^2 - 7n + 14}{8}\right)\nabla^\alpha\xo{q}{\phi}\nabla_\alpha\xo{q}{\phi} - \frac{(n-2)(n-5)}{4}\nabla^\mu\xo{q}{\phi}\nabla_\nu\xo{q}{\phi}\right] \nonumber \\
+\kappa\sum_q\xo{q}{\sigma}\xo{q}{\exp}(\phi\sigma)\left[\delta^\mu_\nu\nabla^\alpha\nabla_\alpha\xo{q}{\phi} - \nabla^\mu\nabla_\nu\xo{q}{\phi}\right] + \kappa\sum_q\xo{q}{\exp}(2\phi\sigma)\left[V(\xo{q}{\phi})\delta^\mu_\nu - 8\pi G\xo{q}{T^\mu_\nu}\right] - 8\pi G \xo{p}{T^\mu_\nu} + O(\kappa^2) = 0
\end{align}
while the scalar equation of motion is
\begin{align}
\xo{j}{\exp}(\sigma\phi)&\left[\xo{j}{\sigma}\xo{p}{R} + \frac{(n-1)(n-6)}{4}\left(2\xo{j}{\sigma^2}\nabla^\alpha\nabla_\alpha\xo{j}{\phi} + \xo{j}{\sigma^3}\nabla^\alpha\xo{j}{\phi}\nabla_\alpha\xo{j}{\phi}\right)\right] = \nonumber \\
&2\left[\frac{\xo{j}{\partial V}}{\partial \phi}\delta^p_j + (1-\delta^p_j) \xo{j}{\exp}(2\sigma\phi)\left(2\xo{j}{\sigma} \xo{j}{V} + \frac{\xo{j}{\partial V}}{\partial \phi}\right)\right] - \xo{j}{\exp}(2\sigma\phi)\left[8\pi G\xo{j}{T}\xo{j}{\sigma}\right] + O(\kappa) = 0
\end{align}
The branching conservation constraint in $n$-dimensions takes the following form
\begin{align}
\sigma\nabla_\mu\phi\Lambda^\mu_\nu + \nabla_\mu\Lambda^\mu_\nu =\Bigg|_{t_*} \exp(\phi \sigma)\left[8\pi G \xo{p}{\nabla}_\mu\xo{q}{T^\mu_\nu} - \nabla_\nu V + 2\sigma\nabla_\mu\phi\left(8\pi G\xo{q}{T^\mu_\nu} - V\delta^\mu_\nu\right)\right]
\label{eqn:branching_constraint}
\end{align}
where
\begin{align}
\Lambda^\mu_\nu \equiv G^\mu_\nu + \sigma^2\delta^\mu_\nu \left(\frac{n^2-7n+14}{8}\right)\nabla^\alpha\phi\nabla_\alpha\phi - \frac{(n-2)(n-5)}{4}\sigma^2\nabla^\mu\phi\nabla_\nu\phi + \sigma(\delta^\mu_\nu\nabla^\alpha\nabla_\alpha\phi - \nabla^\mu\nabla_\nu\phi).
\label{eqn:Lambda}
\end{align}
\end{appendix}

\bibliography{amiec}

\begin{thebibliography}{27}%
\makeatletter
\providecommand \@ifxundefined [1]{%
 \@ifx{#1\undefined}
}%
\providecommand \@ifnum [1]{%
 \ifnum #1\expandafter \@firstoftwo
 \else \expandafter \@secondoftwo
 \fi
}%
\providecommand \@ifx [1]{%
 \ifx #1\expandafter \@firstoftwo
 \else \expandafter \@secondoftwo
 \fi
}%
\providecommand \natexlab [1]{#1}%
\providecommand \enquote  [1]{``#1''}%
\providecommand \bibnamefont  [1]{#1}%
\providecommand \bibfnamefont [1]{#1}%
\providecommand \citenamefont [1]{#1}%
\providecommand \href@noop [0]{\@secondoftwo}%
\providecommand \href [0]{\begingroup \@sanitize@url \@href}%
\providecommand \@href[1]{\@@startlink{#1}\@@href}%
\providecommand \@@href[1]{\endgroup#1\@@endlink}%
\providecommand \@sanitize@url [0]{\catcode `\\12\catcode `\$12\catcode
  `\&12\catcode `\#12\catcode `\^12\catcode `\_12\catcode `\%12\relax}%
\providecommand \@@startlink[1]{}%
\providecommand \@@endlink[0]{}%
\providecommand \url  [0]{\begingroup\@sanitize@url \@url }%
\providecommand \@url [1]{\endgroup\@href {#1}{\urlprefix }}%
\providecommand \urlprefix  [0]{URL }%
\providecommand \Eprint [0]{\href }%
\providecommand \doibase [0]{http://dx.doi.org/}%
\providecommand \selectlanguage [0]{\@gobble}%
\providecommand \bibinfo  [0]{\@secondoftwo}%
\providecommand \bibfield  [0]{\@secondoftwo}%
\providecommand \translation [1]{[#1]}%
\providecommand \BibitemOpen [0]{}%
\providecommand \bibitemStop [0]{}%
\providecommand \bibitemNoStop [0]{.\EOS\space}%
\providecommand \EOS [0]{\spacefactor3000\relax}%
\providecommand \BibitemShut  [1]{\csname bibitem#1\endcsname}%
\let\auto@bib@innerbib\@empty
%</preamble>
\bibitem [{\citenamefont {Smolin}(1992)}]{smolin1992did}%
  \BibitemOpen
  \bibfield  {author} {\bibinfo {author} {\bibfnamefont {L.}~\bibnamefont
  {Smolin}},\ }\href@noop {} {\bibfield  {journal} {\bibinfo  {journal}
  {Classical and Quantum Gravity}\ }\textbf {\bibinfo {volume} {9}},\ \bibinfo
  {pages} {173} (\bibinfo {year} {1992})}\BibitemShut {NoStop}%
\bibitem [{\citenamefont {Smolin}(2006)}]{smolin2006status}%
  \BibitemOpen
  \bibfield  {author} {\bibinfo {author} {\bibfnamefont {L.}~\bibnamefont
  {Smolin}},\ }\href@noop {} {\bibfield  {journal} {\bibinfo  {journal} {arXiv
  preprint hep-th/0612185}\ } (\bibinfo {year} {2006})}\BibitemShut {NoStop}%
\bibitem [{\citenamefont {Rothman}\ and\ \citenamefont
  {Ellis}(1993)}]{rothman1993smolin}%
  \BibitemOpen
  \bibfield  {author} {\bibinfo {author} {\bibfnamefont {T.}~\bibnamefont
  {Rothman}}\ and\ \bibinfo {author} {\bibfnamefont {G.}~\bibnamefont
  {Ellis}},\ }\href@noop {} {\bibfield  {journal} {\bibinfo  {journal}
  {Quarterly Journal of the Royal Astronomical Society}\ }\textbf {\bibinfo
  {volume} {34}},\ \bibinfo {pages} {201} (\bibinfo {year} {1993})}\BibitemShut
  {NoStop}%
\bibitem [{\citenamefont {Darwin}(1872)}]{darwin1872origin}%
  \BibitemOpen
  \bibfield  {author} {\bibinfo {author} {\bibfnamefont {C.}~\bibnamefont
  {Darwin}},\ }\href@noop {} {\emph {\bibinfo {title} {The origin of
  species}}}\ (\bibinfo  {publisher} {Lulu. com},\ \bibinfo {year}
  {1872})\BibitemShut {NoStop}%
\bibitem [{\citenamefont {Damour}\ and\ \citenamefont
  {Esposito-Farese}(1992)}]{damour1992tensor}%
  \BibitemOpen
  \bibfield  {author} {\bibinfo {author} {\bibfnamefont {T.}~\bibnamefont
  {Damour}}\ and\ \bibinfo {author} {\bibfnamefont {G.}~\bibnamefont
  {Esposito-Farese}},\ }\href@noop {} {\bibfield  {journal} {\bibinfo
  {journal} {Classical and Quantum Gravity}\ }\textbf {\bibinfo {volume} {9}},\
  \bibinfo {pages} {2093} (\bibinfo {year} {1992})}\BibitemShut {NoStop}%
\bibitem [{\citenamefont {Damour}\ and\ \citenamefont
  {Esposito-Farese}(1993)}]{damour1993nonperturbative}%
  \BibitemOpen
  \bibfield  {author} {\bibinfo {author} {\bibfnamefont {T.}~\bibnamefont
  {Damour}}\ and\ \bibinfo {author} {\bibfnamefont {G.}~\bibnamefont
  {Esposito-Farese}},\ }\href@noop {} {\bibfield  {journal} {\bibinfo
  {journal} {Physical Review Letters}\ }\textbf {\bibinfo {volume} {70}},\
  \bibinfo {pages} {2220} (\bibinfo {year} {1993})}\BibitemShut {NoStop}%
\bibitem [{\citenamefont {Bassett}\ \emph {et~al.}(2006)\citenamefont
  {Bassett}, \citenamefont {Tsujikawa},\ and\ \citenamefont
  {Wands}}]{bassett2006inflation}%
  \BibitemOpen
  \bibfield  {author} {\bibinfo {author} {\bibfnamefont {B.~A.}\ \bibnamefont
  {Bassett}}, \bibinfo {author} {\bibfnamefont {S.}~\bibnamefont {Tsujikawa}},
  \ and\ \bibinfo {author} {\bibfnamefont {D.}~\bibnamefont {Wands}},\
  }\href@noop {} {\bibfield  {journal} {\bibinfo  {journal} {Reviews of Modern
  Physics}\ }\textbf {\bibinfo {volume} {78}},\ \bibinfo {pages} {537}
  (\bibinfo {year} {2006})}\BibitemShut {NoStop}%
\bibitem [{\citenamefont {Rosen}(1963)}]{rosen1963flat}%
  \BibitemOpen
  \bibfield  {author} {\bibinfo {author} {\bibfnamefont {N.}~\bibnamefont
  {Rosen}},\ }\href@noop {} {\bibfield  {journal} {\bibinfo  {journal} {Annals
  of Physics}\ }\textbf {\bibinfo {volume} {22}},\ \bibinfo {pages} {1}
  (\bibinfo {year} {1963})}\BibitemShut {NoStop}%
\bibitem [{\citenamefont {{Will}}(1993)}]{TEGP}%
  \BibitemOpen
  \bibfield  {author} {\bibinfo {author} {\bibfnamefont {C.~M.}\ \bibnamefont
  {{Will}}},\ }\href@noop {} {\emph {\bibinfo {title} {{Theory and Experiment
  in Gravitational Physics}}}}\ (\bibinfo  {publisher} {Cambridge University
  Press},\ \bibinfo {year} {1993})\BibitemShut {NoStop}%
\bibitem [{\citenamefont {Boulanger}\ \emph {et~al.}(2001)\citenamefont
  {Boulanger}, \citenamefont {Damour}, \citenamefont {Gualtieri},\ and\
  \citenamefont {Henneaux}}]{boulanger2001inconsistency}%
  \BibitemOpen
  \bibfield  {author} {\bibinfo {author} {\bibfnamefont {N.}~\bibnamefont
  {Boulanger}}, \bibinfo {author} {\bibfnamefont {T.}~\bibnamefont {Damour}},
  \bibinfo {author} {\bibfnamefont {L.}~\bibnamefont {Gualtieri}}, \ and\
  \bibinfo {author} {\bibfnamefont {M.}~\bibnamefont {Henneaux}},\ }\href@noop
  {} {\bibfield  {journal} {\bibinfo  {journal} {Nuclear Physics B}\ }\textbf
  {\bibinfo {volume} {597}},\ \bibinfo {pages} {127} (\bibinfo {year}
  {2001})}\BibitemShut {NoStop}%
\bibitem [{\citenamefont {Hassan}\ and\ \citenamefont
  {Rosen}(2012)}]{hassan2012bimetric}%
  \BibitemOpen
  \bibfield  {author} {\bibinfo {author} {\bibfnamefont {S.~F.}\ \bibnamefont
  {Hassan}}\ and\ \bibinfo {author} {\bibfnamefont {R.~A.}\ \bibnamefont
  {Rosen}},\ }\href@noop {} {\bibfield  {journal} {\bibinfo  {journal} {Journal
  of High Energy Physics}\ }\textbf {\bibinfo {volume} {2012}},\ \bibinfo
  {pages} {1} (\bibinfo {year} {2012})}\BibitemShut {NoStop}%
\bibitem [{\citenamefont {O'neill}(1983)}]{o1983semi}%
  \BibitemOpen
  \bibfield  {author} {\bibinfo {author} {\bibfnamefont {B.}~\bibnamefont
  {O'neill}},\ }\href@noop {} {\emph {\bibinfo {title} {Semi-Riemannian
  Geometry With Applications to Relativity, 103}}},\ Vol.\ \bibinfo {volume}
  {103}\ (\bibinfo  {publisher} {Academic press},\ \bibinfo {year}
  {1983})\BibitemShut {NoStop}%
\bibitem [{\citenamefont {Fock}(2015)}]{fock2015theory}%
  \BibitemOpen
  \bibfield  {author} {\bibinfo {author} {\bibfnamefont {V.}~\bibnamefont
  {Fock}},\ }\href@noop {} {\emph {\bibinfo {title} {The theory of space, time
  and gravitation}}}\ (\bibinfo  {publisher} {Elsevier},\ \bibinfo {year}
  {2015})\BibitemShut {NoStop}%
\bibitem [{\citenamefont {Hohmann}\ and\ \citenamefont
  {Wohlfarth}(2010)}]{hohmann2010repulsive}%
  \BibitemOpen
  \bibfield  {author} {\bibinfo {author} {\bibfnamefont {M.}~\bibnamefont
  {Hohmann}}\ and\ \bibinfo {author} {\bibfnamefont {M.~N.}\ \bibnamefont
  {Wohlfarth}},\ }\href@noop {} {\bibfield  {journal} {\bibinfo  {journal}
  {Physical Review D}\ }\textbf {\bibinfo {volume} {81}},\ \bibinfo {pages}
  {104006} (\bibinfo {year} {2010})}\BibitemShut {NoStop}%
\bibitem [{\citenamefont {Chapline}(2003)}]{chapline2003quantum}%
  \BibitemOpen
  \bibfield  {author} {\bibinfo {author} {\bibfnamefont {G.}~\bibnamefont
  {Chapline}},\ }\href@noop {} {\bibfield  {journal} {\bibinfo  {journal}
  {International Journal of Modern Physics A}\ }\textbf {\bibinfo {volume}
  {18}},\ \bibinfo {pages} {3587} (\bibinfo {year} {2003})}\BibitemShut
  {NoStop}%
\bibitem [{\citenamefont {Chapline}\ \emph {et~al.}(2001)\citenamefont
  {Chapline}, \citenamefont {Hohlfeld}, \citenamefont {Laughlin},\ and\
  \citenamefont {Santiago}}]{chapline2001quantum}%
  \BibitemOpen
  \bibfield  {author} {\bibinfo {author} {\bibfnamefont {G.}~\bibnamefont
  {Chapline}}, \bibinfo {author} {\bibfnamefont {E.}~\bibnamefont {Hohlfeld}},
  \bibinfo {author} {\bibfnamefont {R.}~\bibnamefont {Laughlin}}, \ and\
  \bibinfo {author} {\bibfnamefont {D.}~\bibnamefont {Santiago}},\ }\href@noop
  {} {\bibfield  {journal} {\bibinfo  {journal} {Philosophical Magazine Part
  B}\ }\textbf {\bibinfo {volume} {81}},\ \bibinfo {pages} {235} (\bibinfo
  {year} {2001})}\BibitemShut {NoStop}%
\bibitem [{\citenamefont {Rosen}\ and\ \citenamefont
  {Krithivasan}(1995)}]{rosen1995discrete}%
  \BibitemOpen
  \bibfield  {author} {\bibinfo {author} {\bibfnamefont {K.~H.}\ \bibnamefont
  {Rosen}}\ and\ \bibinfo {author} {\bibfnamefont {K.}~\bibnamefont
  {Krithivasan}},\ }\href@noop {} {\emph {\bibinfo {title} {Discrete
  mathematics and its applications}}},\ Vol.~\bibinfo {volume} {6}\ (\bibinfo
  {publisher} {McGraw-Hill New York},\ \bibinfo {year} {1995})\BibitemShut
  {NoStop}%
\bibitem [{\citenamefont {Abbott}\ \emph {et~al.}(2016)\citenamefont {Abbott},
  \citenamefont {Abbott}, \citenamefont {Abbott}, \citenamefont {Abernathy},
  \citenamefont {Acernese}, \citenamefont {Ackley}, \citenamefont {Adams},
  \citenamefont {Adams}, \citenamefont {Addesso}, \citenamefont {Adhikari}
  \emph {et~al.}}]{abbott2016gw151226}%
  \BibitemOpen
  \bibfield  {author} {\bibinfo {author} {\bibfnamefont {B.}~\bibnamefont
  {Abbott}}, \bibinfo {author} {\bibfnamefont {R.}~\bibnamefont {Abbott}},
  \bibinfo {author} {\bibfnamefont {T.}~\bibnamefont {Abbott}}, \bibinfo
  {author} {\bibfnamefont {M.}~\bibnamefont {Abernathy}}, \bibinfo {author}
  {\bibfnamefont {F.}~\bibnamefont {Acernese}}, \bibinfo {author}
  {\bibfnamefont {K.}~\bibnamefont {Ackley}}, \bibinfo {author} {\bibfnamefont
  {C.}~\bibnamefont {Adams}}, \bibinfo {author} {\bibfnamefont
  {T.}~\bibnamefont {Adams}}, \bibinfo {author} {\bibfnamefont
  {P.}~\bibnamefont {Addesso}}, \bibinfo {author} {\bibfnamefont
  {R.}~\bibnamefont {Adhikari}},  \emph {et~al.},\ }\href@noop {} {\bibfield
  {journal} {\bibinfo  {journal} {Physical Review Letters}\ }\textbf {\bibinfo
  {volume} {116}},\ \bibinfo {pages} {241103} (\bibinfo {year}
  {2016})}\BibitemShut {NoStop}%
\bibitem [{\citenamefont {Carroll}(2004)}]{carroll2004spacetime}%
  \BibitemOpen
  \bibfield  {author} {\bibinfo {author} {\bibfnamefont {S.~M.}\ \bibnamefont
  {Carroll}},\ }\href@noop {} {\emph {\bibinfo {title} {Spacetime and geometry.
  An introduction to general relativity}}},\ Vol.~\bibinfo {volume} {1}\
  (\bibinfo {year} {2004})\BibitemShut {NoStop}%
\bibitem [{\citenamefont {Kuusk}\ \emph {et~al.}(2016)\citenamefont {Kuusk},
  \citenamefont {J{\"a}rv},\ and\ \citenamefont {Vilson}}]{kuusk2016invariant}%
  \BibitemOpen
  \bibfield  {author} {\bibinfo {author} {\bibfnamefont {P.}~\bibnamefont
  {Kuusk}}, \bibinfo {author} {\bibfnamefont {L.}~\bibnamefont {J{\"a}rv}}, \
  and\ \bibinfo {author} {\bibfnamefont {O.}~\bibnamefont {Vilson}},\
  }\href@noop {} {\bibfield  {journal} {\bibinfo  {journal} {International
  Journal of Modern Physics A}\ }\textbf {\bibinfo {volume} {31}},\ \bibinfo
  {pages} {1641003} (\bibinfo {year} {2016})}\BibitemShut {NoStop}%
\bibitem [{\citenamefont {Heymans}\ \emph {et~al.}(2012)\citenamefont
  {Heymans}, \citenamefont {Van~Waerbeke}, \citenamefont {Miller},
  \citenamefont {Erben}, \citenamefont {Hildebrandt}, \citenamefont {Hoekstra},
  \citenamefont {Kitching}, \citenamefont {Mellier}, \citenamefont {Simon},
  \citenamefont {Bonnett} \emph {et~al.}}]{heymans2012cfhtlens}%
  \BibitemOpen
  \bibfield  {author} {\bibinfo {author} {\bibfnamefont {C.}~\bibnamefont
  {Heymans}}, \bibinfo {author} {\bibfnamefont {L.}~\bibnamefont
  {Van~Waerbeke}}, \bibinfo {author} {\bibfnamefont {L.}~\bibnamefont
  {Miller}}, \bibinfo {author} {\bibfnamefont {T.}~\bibnamefont {Erben}},
  \bibinfo {author} {\bibfnamefont {H.}~\bibnamefont {Hildebrandt}}, \bibinfo
  {author} {\bibfnamefont {H.}~\bibnamefont {Hoekstra}}, \bibinfo {author}
  {\bibfnamefont {T.~D.}\ \bibnamefont {Kitching}}, \bibinfo {author}
  {\bibfnamefont {Y.}~\bibnamefont {Mellier}}, \bibinfo {author} {\bibfnamefont
  {P.}~\bibnamefont {Simon}}, \bibinfo {author} {\bibfnamefont
  {C.}~\bibnamefont {Bonnett}},  \emph {et~al.},\ }\href@noop {} {\bibfield
  {journal} {\bibinfo  {journal} {Monthly Notices of the Royal Astronomical
  Society}\ }\textbf {\bibinfo {volume} {427}},\ \bibinfo {pages} {146}
  (\bibinfo {year} {2012})}\BibitemShut {NoStop}%
\bibitem [{\citenamefont {Hohmann}\ \emph {et~al.}(2016)\citenamefont
  {Hohmann}, \citenamefont {Jarv}, \citenamefont {Kuusk}, \citenamefont
  {Randla},\ and\ \citenamefont {Vilson}}]{HohmannErik}%
  \BibitemOpen
  \bibfield  {author} {\bibinfo {author} {\bibfnamefont {M.}~\bibnamefont
  {Hohmann}}, \bibinfo {author} {\bibfnamefont {L.}~\bibnamefont {Jarv}},
  \bibinfo {author} {\bibfnamefont {P.}~\bibnamefont {Kuusk}}, \bibinfo
  {author} {\bibfnamefont {E.}~\bibnamefont {Randla}}, \ and\ \bibinfo {author}
  {\bibfnamefont {O.}~\bibnamefont {Vilson}},\ }\href@noop {} {\  (\bibinfo
  {year} {2016})},\ \Eprint {http://arxiv.org/abs/1607.02356} {arXiv:1607.02356
  [gr-qc]} \BibitemShut {NoStop}%
%%CITATION = ARXIV:1607.02356;%%
\bibitem [{\citenamefont {Hohmann}(2014)}]{HohmannPPN}%
  \BibitemOpen
  \bibfield  {author} {\bibinfo {author} {\bibfnamefont {M.}~\bibnamefont
  {Hohmann}},\ }\href {\doibase 10.1088/0264-9381/31/13/135003} {\bibfield
  {journal} {\bibinfo  {journal} {Class. Quant. Grav.}\ }\textbf {\bibinfo
  {volume} {31}},\ \bibinfo {pages} {135003} (\bibinfo {year} {2014})},\
  \Eprint {http://arxiv.org/abs/1309.7787} {arXiv:1309.7787 [gr-qc]}
  \BibitemShut {NoStop}%
%%CITATION = ARXIV:1309.7787;%%
\bibitem [{\citenamefont {Mahdavi}\ \emph {et~al.}(2007)\citenamefont
  {Mahdavi}, \citenamefont {Hoekstra}, \citenamefont {Babul}, \citenamefont
  {Balam},\ and\ \citenamefont {Capak}}]{mahdavi2007dark}%
  \BibitemOpen
  \bibfield  {author} {\bibinfo {author} {\bibfnamefont {A.}~\bibnamefont
  {Mahdavi}}, \bibinfo {author} {\bibfnamefont {H.}~\bibnamefont {Hoekstra}},
  \bibinfo {author} {\bibfnamefont {A.}~\bibnamefont {Babul}}, \bibinfo
  {author} {\bibfnamefont {D.~D.}\ \bibnamefont {Balam}}, \ and\ \bibinfo
  {author} {\bibfnamefont {P.~L.}\ \bibnamefont {Capak}},\ }\href@noop {}
  {\bibfield  {journal} {\bibinfo  {journal} {The Astrophysical Journal}\
  }\textbf {\bibinfo {volume} {668}},\ \bibinfo {pages} {806} (\bibinfo {year}
  {2007})}\BibitemShut {NoStop}%
\bibitem [{\citenamefont {Clowe}\ \emph {et~al.}(2006)\citenamefont {Clowe},
  \citenamefont {Brada{\v{c}}}, \citenamefont {Gonzalez}, \citenamefont
  {Markevitch}, \citenamefont {Randall}, \citenamefont {Jones},\ and\
  \citenamefont {Zaritsky}}]{clowe2006direct}%
  \BibitemOpen
  \bibfield  {author} {\bibinfo {author} {\bibfnamefont {D.}~\bibnamefont
  {Clowe}}, \bibinfo {author} {\bibfnamefont {M.}~\bibnamefont {Brada{\v{c}}}},
  \bibinfo {author} {\bibfnamefont {A.~H.}\ \bibnamefont {Gonzalez}}, \bibinfo
  {author} {\bibfnamefont {M.}~\bibnamefont {Markevitch}}, \bibinfo {author}
  {\bibfnamefont {S.~W.}\ \bibnamefont {Randall}}, \bibinfo {author}
  {\bibfnamefont {C.}~\bibnamefont {Jones}}, \ and\ \bibinfo {author}
  {\bibfnamefont {D.}~\bibnamefont {Zaritsky}},\ }\href@noop {} {\bibfield
  {journal} {\bibinfo  {journal} {The Astrophysical Journal Letters}\ }\textbf
  {\bibinfo {volume} {648}},\ \bibinfo {pages} {L109} (\bibinfo {year}
  {2006})}\BibitemShut {NoStop}%
\bibitem [{\citenamefont {Milgrom}(1983)}]{milgrom1983modification}%
  \BibitemOpen
  \bibfield  {author} {\bibinfo {author} {\bibfnamefont {M.}~\bibnamefont
  {Milgrom}},\ }\href@noop {} {\bibfield  {journal} {\bibinfo  {journal} {The
  Astrophysical Journal}\ }\textbf {\bibinfo {volume} {270}},\ \bibinfo {pages}
  {365} (\bibinfo {year} {1983})}\BibitemShut {NoStop}%
\bibitem [{\citenamefont {Bekenstein}(2004)}]{bekenstein2004relativistic}%
  \BibitemOpen
  \bibfield  {author} {\bibinfo {author} {\bibfnamefont {J.~D.}\ \bibnamefont
  {Bekenstein}},\ }\href@noop {} {\bibfield  {journal} {\bibinfo  {journal}
  {Physical Review D}\ }\textbf {\bibinfo {volume} {70}},\ \bibinfo {pages}
  {083509} (\bibinfo {year} {2004})}\BibitemShut {NoStop}%
\end{thebibliography}%

\end{document}